\documentclass[12pt]{article}
\usepackage{pdproc,epsfig,graphics} 
\begin{document}

\newcommand{\lsim}   {\mathrel{\mathop{\kern 0pt \rlap
  {\raise.2ex\hbox{$<$}}}
  \lower.9ex\hbox{\kern-.190em $\sim$}}}
\newcommand{\gsim}   {\mathrel{\mathop{\kern 0pt \rlap
  {\raise.2ex\hbox{$>$}}}
\lower.9ex\hbox{\kern-.190em $\sim$}}}
\def\be{\begin{equation}}
\def\ee{\end{equation}}
\def\ba{\begin{eqnarray}}
\def\ea{\end{eqnarray}}
\def\d{{\rm d}}
\def\ap{\approx}

\title{
UHE NEUTRINO ASTRONOMY AND NEUTRINO OSCILLATIONS}

\author{V. Berezinsky}

\address{INFN, Laboratori Nazionali del Gran Sasso,\\ 
67010 Assergi (AQ) Italy}

\abstract{ 
UHE neutrinos with  $E>10^{17}$~eV can be produced by ultra-high energy 
cosmic rays (UHECR) interacting with CMB photons  ({\em cosmogenic neutrinos}) 
and by top-down sources, such as  topological defects (TD), superheavy
dark matter (SHDM) and mirror matter. Cosmogenic neutrinos are
reliably predicted and their fluxes can be numerically evaluated 
using the observed flux of UHECR. The lower limit for the flux is obtained 
for the case of pure proton composition of the observed UHECR. The rigorous 
upper limit for cosmogenic neutrino flux also exists. The maximum neutrino 
energy is determined by maximum energy of acceleration, which at
least for the shock acceleration is expected not to exceed 
$10^{21} - 10^{22}$~eV. The top-down sources provide neutrino energies 
a few orders of magnitude higher, and this can be considered as a
signature of these models. Oscillations play important role in UHE 
neutrino astronomy. At production of cosmogenic neutrinos 
$\tau$-neutrinos are absent and $\bar{\nu}_e$ neutrinos are
suppressed. These species, important for detection, appear in the
observed fluxes due to oscillation. Mirror neutrinos cannot be
observed directly, but due to oscillations to ordinary neutrinos they 
can provide the largest neutrino flux at the highest energies.
}

\normalsize\baselineskip=15pt

\section{Introduction}

The boundary $E_{\nu} \sim 10^{17}$~eV between HE and UHE neutrino astronomy 
is connected with observational technique. The
neutrino observations with underwater/ice detectors are valid mostly
for HE neutrino range. This is successful and well developed
experimental technique.   
Two detectors, Baikal and AMANDA, has reached now the saturated 
statistics, two other arrays, IceCube and ANTARES, started the
observations, and two  projects, NESTOR and NEMO, are at the stage 
of testing. 

The prospects for UHE neutrino astronomy appeared in 1960s soon after 
prediction of  the GZK \cite{GZK} cutoff. It has
been realized \cite{BZ} that proton interaction with CMB photons
at large redshifts in case of cosmological evolution of the sources 
can produce UHE neutrino fluxes much higher than the observed UHECR 
flux.

In 80s it was understood that topological defects can
produce unstable superheavy particles with
masses up to the  GUT scale \cite{Witten} and neutrinos with
tremendous energies  can emerge due to this process \cite{HiSch}. 

It has been proposed that UHE 
neutrinos can be detected observing the horizontal Extensive Air Showers  
(EAS) \cite{BS}. 
The exciting prospects for detection of UHE neutrinos have appeared 
with the ideas of space detection, e.g. in the projects 
EUSO\cite{EUSO} and OWL\cite{OWL}. At present there is well developed 
JEM-EUSO project \cite{jemeuso} with the prospects to start the
observations in 2012 - 2013.
 
The basic idea of detection by EUSO is similar to the fluorescent technique 
for observations of extensive air showers (EAS) from the surface of 
the Earth. The UHE neutrino entering the Earth atmosphere  
produces an EAS. The known fraction of its energy, which
reaches 90\% , is radiated in form of isotropic fluorescent light. 
An optical telescope from the space detects it.
Since the observatory is located at very large height 
($\sim 400$~km) in comparison with thickness of the atmosphere, the
fraction of detected flux is known, and thus this is the calorimetric 
measurement (absorption of up-going photons is 
small). A telescope with diameter 2.5 m controls the area 
$\sim 10^5$~km$^2$ and has a threshold for EAS detection 
$E_{\rm th}\sim 1\times 10^{19}$~eV \cite{jemeuso}.    	 

The very efficient method of UHE neutrino detection is given by 
observations of radio emission by neutrino-induced showers in ice 
and lunar regolith.
This method has been originally suggested by G.~Askaryan 
in 60s \cite{askarian}. Propagating in the matter the shower acquires 
excessive negative electric charge due to involvement of the 
matter electrons in knock-on process. The coherent Cerenkov radiation
of these electrons produces the radio pulse. Recently this method has
been confirmed by the laboratory measurements \cite{saltz}. There were 
the experiments to search for such radiation from
neutrino-induced showers in the Greenland and  Antarctic ice and in the
lunar regolith. In all cases the radio-emission can be observed only 
for neutrinos of extremely high energies.
The upper limits on the flux of these neutrinos 
have been obtained: in GLUE experiment \cite{glue} by radiation from
the moon, in FORTE experiment \cite{forte} by radiation from the
Greenland ice and in ANITA \cite{anita} and RICE \cite{rice}
experiments  from the Antarctic ice.

The characteristic feature of these detection methods is 
the high energy threshold
$E \gsim 1\times 10^{19} - 1\times 10^{20}$~eV. How neutrinos of these 
energies can be produced?      

The most conservative mechanism of UHE neutrino production is
$p\gamma$ mechanism of collisions of accelerated protons/nuclei with 
low-energy CMB photons. To provide neutrinos with energies higher that 
$1\times 10^{20}$~eV the accelerated protons must have energies higher 
than $2\times 10^{21}$~eV. For non-relativistic shock 
acceleration this energy can reach optimistically $1\times10^{21}$~eV. 
For relativistic shock this energy can be somewhat higher. 
To have neutrinos with higher energies one has to put his hopes on less 
developed ideas of acceleration such as 
acceleration in strong e-m waves, exotic plasma mechanisms of
acceleration and unipolar induction. 

The top-down scenarios can naturally provide neutrinos with energies
higher and much higher than $1\times 10^{20}$~eV. The idea common  
for many mechanisms is given by existence of superheavy particles with 
very large masses up to GUT scale. In Grand Unified Theories (GUT)
these particles (gauge bosons and higgses) are short-lived. In the
cosmic space they are produced by Topological Defects (TD). The
decay of these particles results in the parton cascade, which is 
terminated by production of pions and other hadrons.
Neutrinos are produced in their decays.

The superheavy particles are naturally produced at post-inflationary 
stage of the universe. The most reliable mechanism of production 
is gravitational one. The masses of the produced particles can reach 
$10^{13} - 10^{14}$~GeV. Protecting by some symmetry (e.g. gauge
symmetry or discrete gauge symmetry like R-parity in supersymmetry),
these particles can survive until present time and
produce neutrinos in the decays or annihilation. 

%%%%%%%%%%%%%%%%%%%%%%%%%%%%%%%%%%%%%%%%%%%%%%%%%%%%%%%%%%%%%%%%%%%%%%%%%
\section{Effects of UHE neutrino oscillations} 
\label{sec:oscillation}
Characteristic distances to the sources of UHE neutrinos are 
much larger than maximum oscillation lengths. Taken as the maximum
neutrino energy $E_{\rm max} \sim 10^{20}$~eV one obtains the maximum 
oscillation length of order of 100~pc
\be
\ell_{\rm osc}^{\rm max}=\frac{4\pi E_{\rm max}}{(\Delta m^2)_{\rm min}}
\approx 120{\rm pc}\left (\frac{E}{10^{20}{\rm eV}}\right )
\left ( \frac{7\times 10^{-5}{\rm eV}^2}{\Delta m^2}\right )
\label{l-osc},
\ee 
while the distances $r$ to UHE neutrino sources even in our galaxy is of
order of a few kpc. The relation $r \gg \ell_{\rm osc}$ is often
interpreted as a sufficient condition for flavour equipartition at observation 
$\nu_e : \nu_\mu : \nu_\tau = 1:1:1$ . However, it
is well known that this is not true for 
arbitrary flavour composition at  generation. In the recent work 
\cite{pakvasa} the precise calculations for connection of the
generation and observed flavour compositions are performed.

The basic idea of calculations in \cite{pakvasa} can be explained in
the simplified form as follows.

Let us consider how observed flavour ratio is connected with  flavour 
composition at  generation in the limiting case $r \gg \ell_{\rm osc}$.
The flavour neutrino eigenstate $\nu_\alpha$ is given by mixing of 
mass eigenstates  $\nu_i$  as $\nu_\alpha = U_{\alpha k} \nu_k$, where 
$\alpha= e,\mu,\tau$,~ are flavour indicies, $k$ are mass eigenstate indicies  
$k=1,2,3$ and  $U_{\alpha k}$  is the mixing matrix. In the most
general case   $U_{\alpha k}$ is expressed through solar $\theta_{12}$ and
atmospheric $\theta_{23}$  neutrino mixing angles, and include also 
small $U_{e3}$ term (see equation below). However with a good accuracy one
may use  $U_{e3}=0$ and  $\theta_{23}=\pi/4$ and thus obtain  
\be 
%\vspace{-2mm}
U = %
\left( %
\begin{array}{ccc}
c_{12} & s_{12} & U_{e3} \\
-s_{12}c_{23} & c_{12}c_{23} & s_{23} \\
s_{12}s_{23} & -c_{12}s_{23} & c_{23} \\
\end{array}
\right) %
\rightarrow
\left( %
\begin{array}{lll}
c_{12} & s_{12} & 0 \\
-\frac{s_{12}}{\sqrt 2} &\frac{c_{12}}{\sqrt 2} &\frac 1{\sqrt 2} \\
\frac{s_{12}}{\sqrt 2} & -\frac{c_{12}}{\sqrt 2} &\frac 1{\sqrt 2}
\end{array}
\right). %
\label{eq:mix-matrix}
\ee
In the limiting case $r \gg \ell_{\rm osc}$,~ $<\sin r/l_{{\rm osc}}>=0$ 
and $<\sin^2 r/l_{{\rm osc}}>=\frac12$,~ and 
the propagation matrix, which describes $\nu_\alpha \to \nu_\beta$ 
oscillation, is given by  
\be
P_{\alpha \beta} = \sum |U_{\alpha i}|^2 |U_{\beta i}|^2
\label{eq:prop-matrix}
\ee 

The propagation matrix $P_{\alpha \beta}$ given by Eq.~(\ref{eq:prop-matrix})
allows to calculate flavour composition at observation for given
flavour composition at generation. In particular for 
normal generation composition valid for unsuppressed channels of $\pi$ decays
$\nu_e: \nu_\mu : \nu_\tau = 1:2:0$ one obtains from
Eq.~(\ref{eq:prop-matrix})     
the composition at observation 1:1:1. 

In the work \cite{pakvasa} are given many examples of absence equipartition  
at observations: the suppression of muon decays in the chain of pion decay,
production of neutrinos in neutron beam, decay of neutrinos and
others. 
%%%%%%%%%%%%%%%%%%%%%%%%%%%%%%%%%%%%%%%%%%%%%%%%%%%%%%%%%%%%%%%%%%%%%%%%%%%%%%
\section{Observational effects caused by oscillations}
\label{sec:obs-oscillation}
In this paper we will consider three observational effects caused by 
oscillations: (i) appearance of $\bar{\nu}_e$ neutrinos important for 
resonant reaction  $ \bar{\nu}_e + e^- 
\rightarrow W^- \rightarrow {\rm hadrons}$,~ (ii) appearance of 
$\tau$ neutrinos and  oscillation of mirror to ordinary neutrinos. 
In this section we consider the first two effects; oscillations
of mirror neutrinos will be discussed in section \ref{sec:mirror}.\\*[2mm]
{\bf Resonant interaction of $\bar{\nu}_e$ neutrinos}.\\*[2mm]
Glashow \cite{Glashow} considered in 1960 the resonance reaction 
 $ \bar{\nu}_e + e^-
\rightarrow W^- \rightarrow \mu^- + \bar{\nu}_\mu$. The resonance
reaction $ \bar{\nu}_e + e^- \rightarrow W^- \rightarrow {\rm  hadrons}$
was first suggested in 1977 in \cite{BG77}. We will follow here the
analytic approach of this paper. 

The resonant production of $W^-$ in $\bar{\nu}_ee$ collisions occurs at
$\bar{\nu}_e$ energy
\be
E_0 = \frac{m_W^2}{2 m_e} = 6.3 \times 10^6 {\rm GeV}.
\label{E-res}
\ee
Integrating over the Breit-Wigner resonance, one obtains analytically
\cite{BG77} the rate of
resonant events in underground detector with number of electrons $N_e$, 
given by   
\be 
\nu_{\rm res} = 2 \pi \sigma_{\rm eff} E_0 J_{\bar{\nu}_e}(E_0) N_e,
\label{res-rate}
\ee
where $J_{\bar{\nu}_e}(E_0)$ is the diffuse flux of UHE  $\bar{\nu}_e$, 
$2\pi$ is the solid angle for which deep underground detector is open
for UHE neutrino flux and $\sigma_{\rm eff}$ is the value left after 
integration over the Breit-Wigner formula, which has meaning of effective
cross-section 
\be 
\sigma_{\rm eff} = \frac{8 \pi}{3 \sqrt{2}} G_F =
2.7\times 10^{-32} {\rm cm}^2,
\label{res-crossection}
\ee
where $G_F$ is the Fermi constant. 

For cosmogenic neutrinos produced in $p\gamma$ collisions with CMB
photons, the production of $\bar{\nu}_e$ is strongly suppressed in 
comparison with $\nu_e$ neutrinos, because the latter are produced 
in the chain of decay of the positively charged pions: 
$\pi^+ \to \mu^+ \to \nu_e$, 
while $\bar{\nu}_e$ in the chain of decay of negatively charged pions
$\pi^- \to \mu^- \to \bar{\nu}_e$. Production of $\pi^+$  occurs 
in the resonant reaction $p+\gamma \to \Delta^+$, while production of 
$\pi^-$ goes  through $p+\gamma \to p + \pi^-+\pi^-$ with a small 
cross-section. As has been first noted in \cite{BG77} 
$\bar{\nu}_\mu \to \nu_e$ oscillation considerably increases the 
flux of cosmogenic $\bar{\nu}_e$ neutrinos. If to take the generation 
flavour ratio $\bar{\nu}_e: \bar{\nu}_\mu : \bar{\nu}_\tau =$0:1:0, 
we obtain from Eq.~(\ref{eq:prop-matrix}) the ratio at observation 
$\bar{\nu}_e: \bar{\nu}_\mu : \bar{\nu}_\tau =\frac{1}{2}:1:1$.
For this flavour ratio the resonant signal is seen well over the
background.  \\
\newpage
\noindent
{\bf Tau neutrinos}\\
Tau neutrinos are not produced in the chain of pion decays and appear 
at observation due to oscillations. 

\begin{figure*}[h]
  \begin{center}
  \mbox{\includegraphics[width=0.7\textwidth]{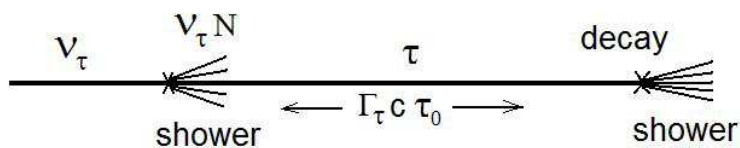}}
  \end{center}
  \caption{ Double-bang effect produced by tau-neutrino in underground 
          detector }
\label{fig:tau}
  \end{figure*}

In underground detectors tau-neutrinos produce the characteristic
two-bang effect \cite{Learned-Pakvasa}: the first hadronic shower
appears in $\nu_{\tau}+N \rightarrow \tau + {\rm hadrons}$
interaction, and the second one is produced by tau-lepton decay 
$\tau \to \nu_{\tau} + {\rm hadrons}$. At energy of tau-neutrino 
$E \sim 10^{15}$~eV the distance between  two showers is about 
50 m, and such event can be observed in IceCube detector. 
The tau lepton propagating between two shower vertex radiates, 
like muon, the Cherenkov photons and this radiation can be also
detected. Fig.~\ref{fig:tau} illustrates the detection of tau-neutrino
in the deep-underground detector.

The UHE neutrinos produce Earth-skimming effect \cite{Fargion}, which 
can be observed by gigantic EAS arrays like the Auger detector.  
\begin{figure*}[h]
  \begin{center}
  \mbox{\includegraphics[width=0.7\textwidth]{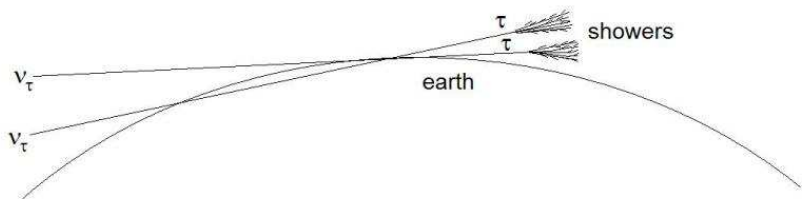}}
  \end{center}
  \caption{ Earth-skimming effect produced by UHE tau-neutrinos, which 
          can be observed by gigantic EAS detectors }
\label{fig:skimming} 
  \end{figure*}
This is a mechanism of detection of horizontally moving neutrinos, but 
tau-neutrinos have the advantages in comparison with other neutrinos.
Tau-neutrino can cross the large thickness of matter in the earth, due 
to effect of regeneration. Interacting with matter it produces tau-lepton, 
which at large energies propagates to large distance and decays
producing again tau-neutrino. At energy $E \gsim 10^{18}$~eV tau-lepton 
has a decay length about 50 km and it can decay in the air producing
hadrons which induce EAS propagating almost parallel to the Earth surface 
(see Fig.~\ref{fig:skimming}). The  surface detectors register the 
EAS electromagnetic component from a lower part of EAS. In this way 
the upper limit on UHE tau-neutrinos was recently obtained at the
Auger detector \cite{tau-auger}.\\    
\newpage
\section{Cosmogenic neutrinos in the dip model}
\label{sec:cosmogenic}
\begin{figure*}[t]
  \begin{center}
  %\mbox{\includegraphics[width=0.7\textwidth]{DoubleBang.eps}}
\mbox{\includegraphics[height=5cm,width=7.5cm]{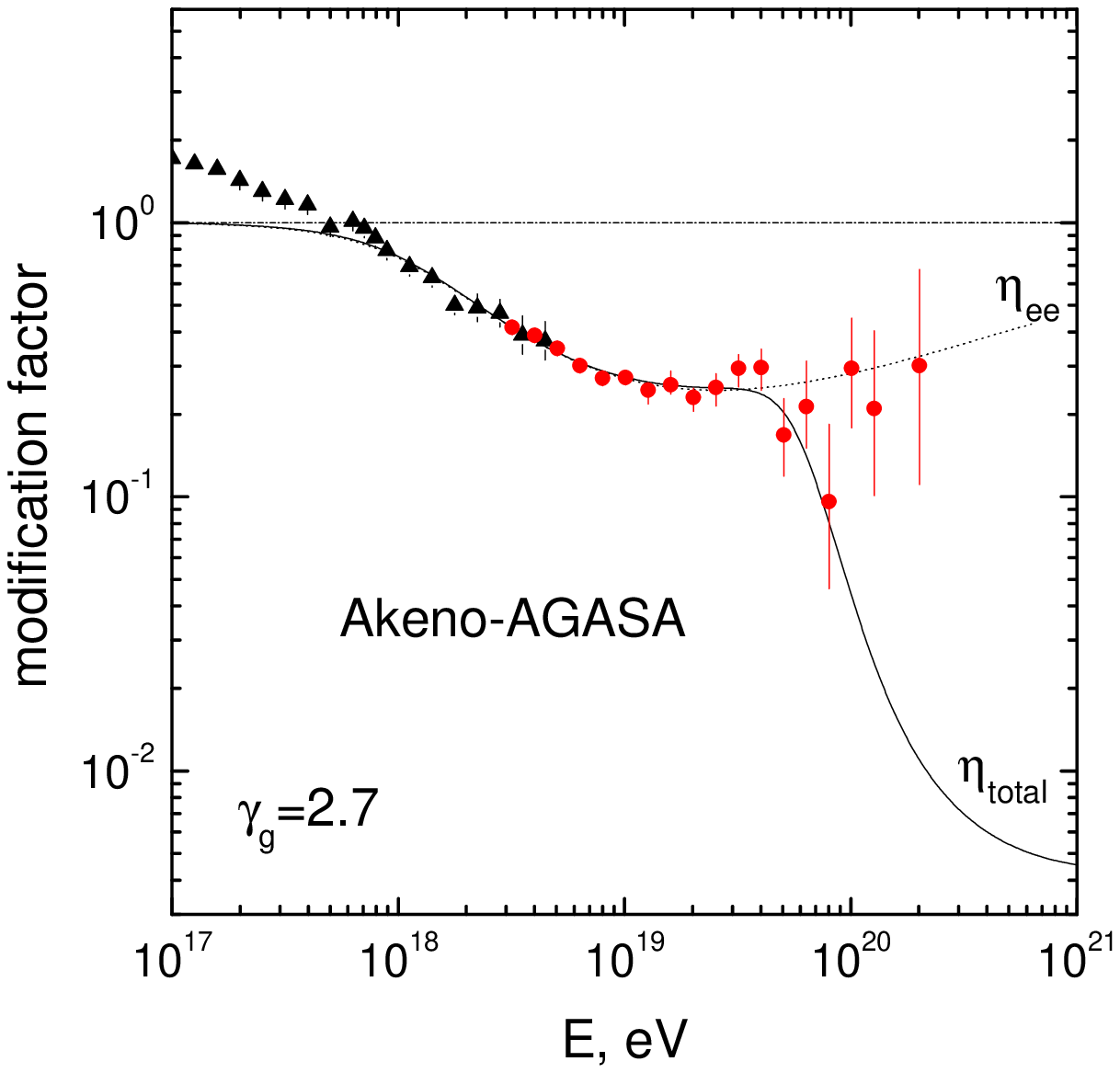}}
\mbox{\includegraphics[height=5cm,width=7.3cm]{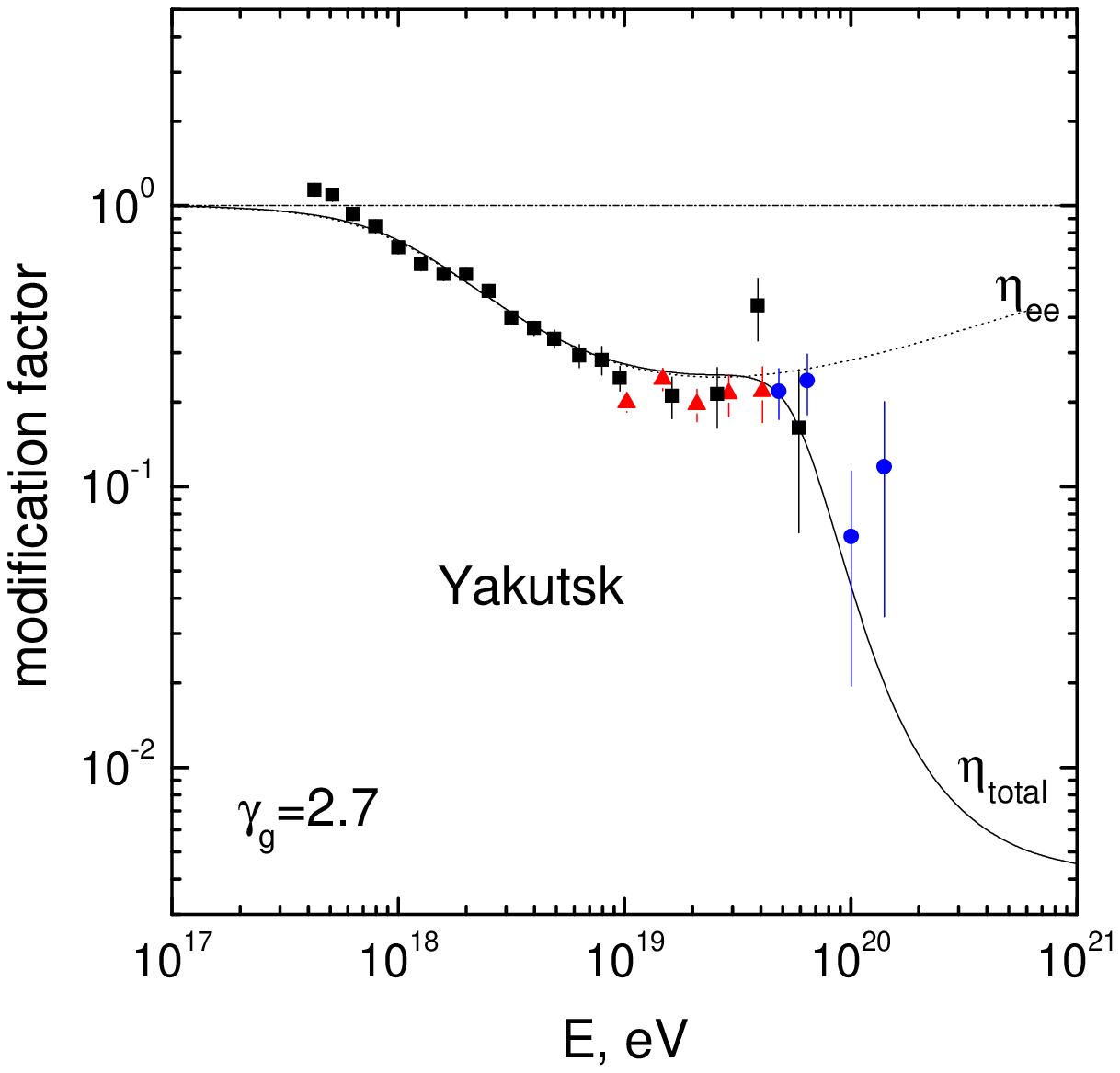}}

\bigskip %\noindent
\mbox{\includegraphics[height=5cm,width=7.5cm]{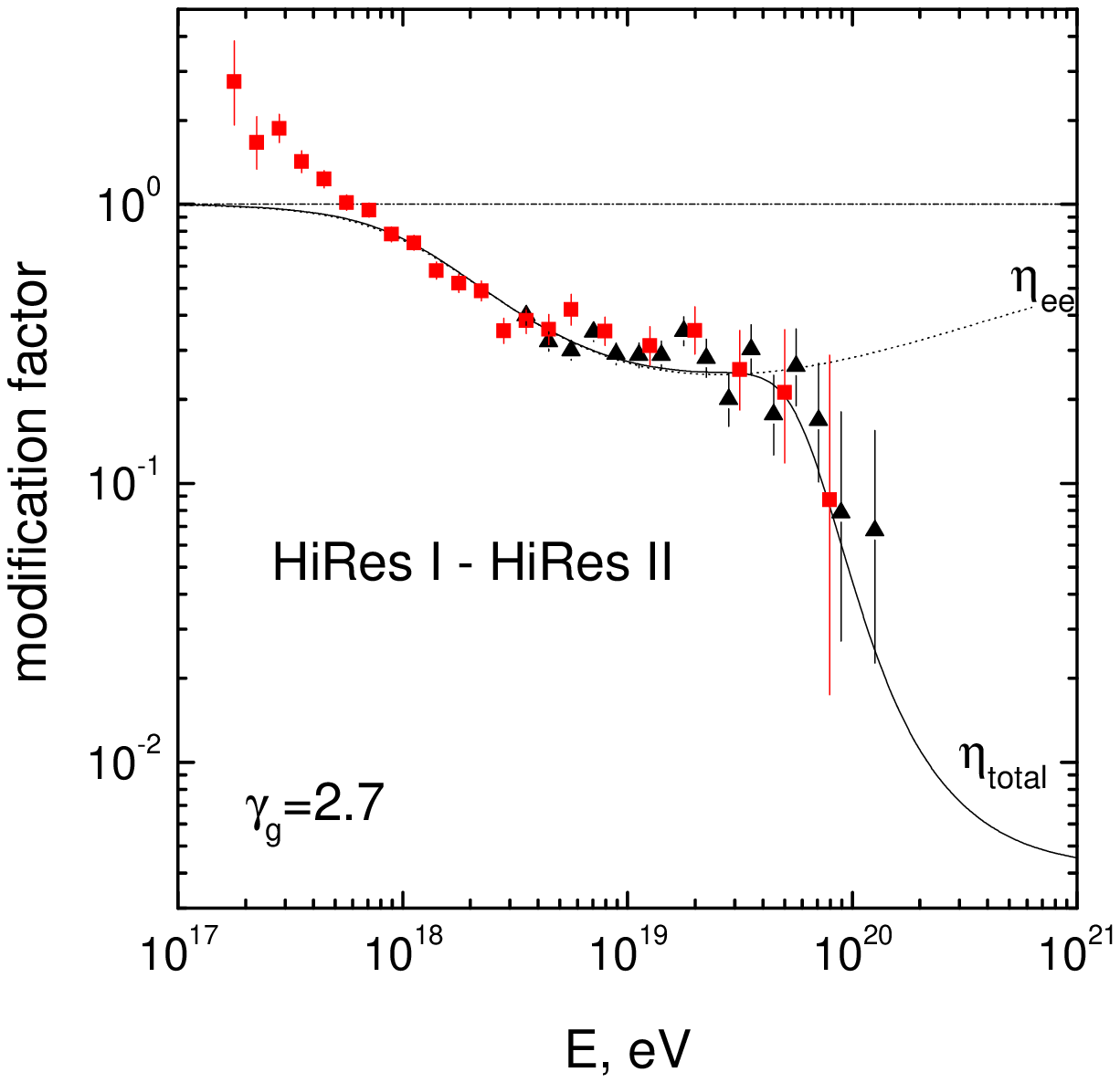}}
\mbox{\includegraphics[height=5cm,width=7.3cm]{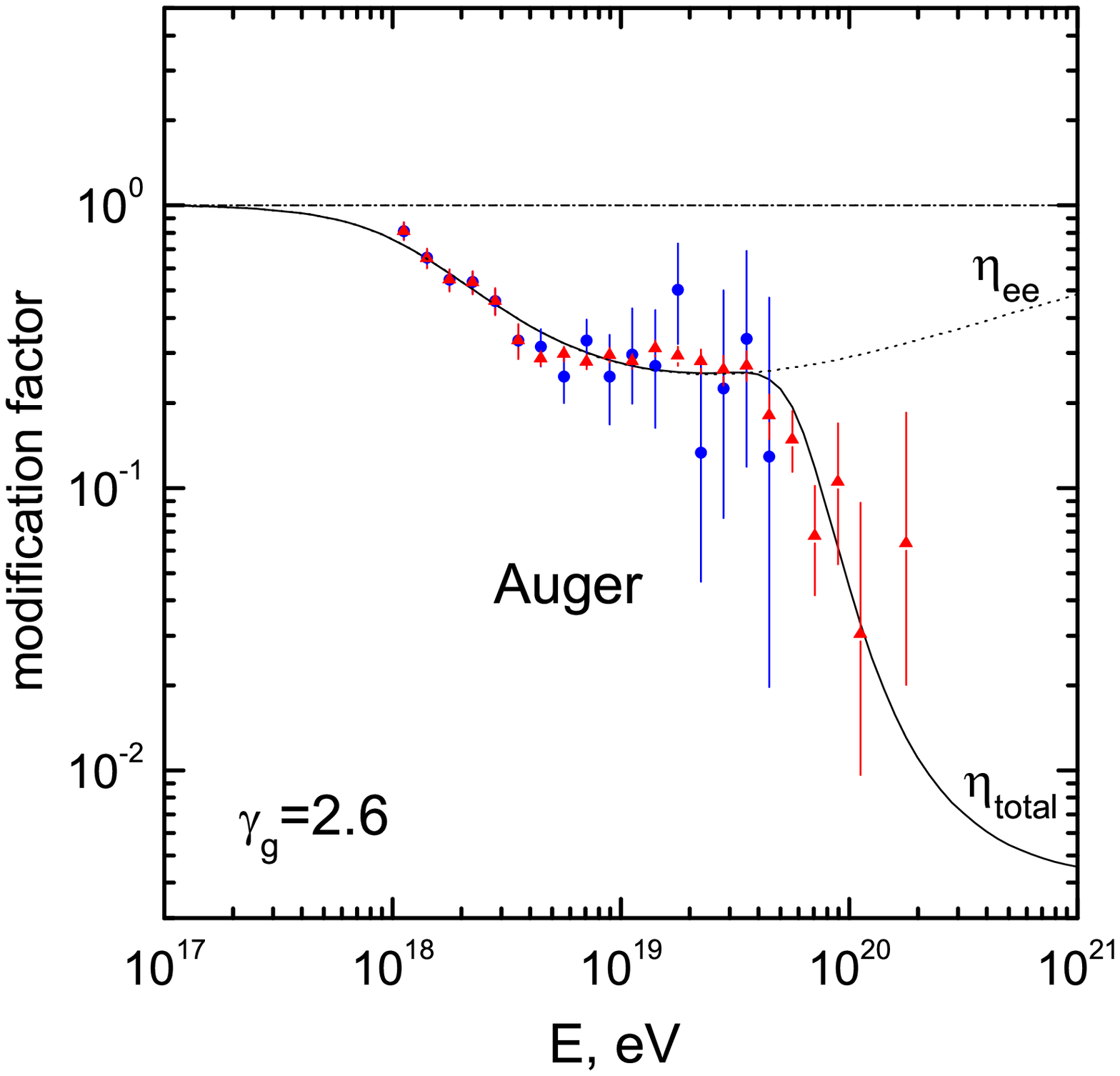}}
\end{center}
\caption{Theoretical pair-production dip and GZK cutoff  in comparison with
the observational data for non-evolutionary models with generation 
index $\gamma_g = 2.6 - 2.7$. The data of HiRes and Auger detectors 
show steepening of the spectrum consistent with the GZK cutoff. The excess
of experimental modification factor over $\eta=1$ at 
$E < 1\times 10^{18}$~ eV evidences for the  new component, which is 
given by galactic cosmic rays.
}
\label{fig:dips}
\end{figure*}

Starting from pioneering work \cite{BZ} the fluxes of cosmogenic
neutrinos have been calculated in many works \cite{ESS} - \cite{Allard}.
The predicted fluxes differ very considerably, depending on the
different assumptions about mass composition of accelerated particles, 
on maximum energy of acceleration and on cosmological 
evolution of the sources. We shall present here the UHE neutrino
fluxes calculated in the dip model for observed 
UHECR \cite{BGG-PL,BGG-prd}.\\*[2mm]                             
{\bf The dip model for UHECR}\\
The pair production dip is a feature of interaction of extragalactic
UHE protons propagating trough CMB radiation. It is caused by energy losses of
protons due to $p+ \gamma_{\rm CMB} \to e^++e^-+p$ scattering. This feature
in proton spectrum, in principle, is very similar to the GZK cutoff
which is caused by photopion production $p+ \gamma_{\rm CMB} \to N+ \pi$.
Both features are convenient to analyse in terms of 
{\em modification factor} $\eta(E)$ , which is defined as a ratio of the 
diffuse 
proton spectrum $J_p(E)$ calculated with all energy losses included
to the so-called unmodified spectrum $J_p^{\rm unm}(E)$, when
only adiabatic energy losses due to expansion of the universe are
taken into account:
\be
\eta(E)=\frac{J_p(E)}{J_p^{\rm unm}(E)},
\label{eq:mfactor}
\ee
The modification factor should be considered as the theoretical 
spectrum. Being defined as a ratio of two spectra it is free from 
many uncertainties. 
The modification factor is calculated using the  power-law generation function 
$Q(E_g) \propto E_g^{-\gamma_g}$, where $E_g$ is the energy of a
proton at generation in a source and $\gamma_g$ is the generation index,
The calculated modification factor gives the excellent agreement with 
observational data for 
$\gamma_g = 2.6 - 2.7$ and for absence of source evolution, i.e. for 
evolutionary factor $(1+z)^m$ with $m=0$. However, inclusion of 
evolution  gives also good agreement with the data. This
is not a surprising thing, since inclusion of evolution means
introducing two additional free parameters, $m$ and $z_{\rm max}$.
In Fig.~\ref{fig:dips} we show 
the comparison of the predicted dip and GZK cutoff with observational
data of Akeno-AGASA, Yakutsk, HiRes and Auger detectors. 

The observable part of the dip is extended from $1\times 10^{18}$~eV,
where modification factor reaches 1, 
and up to $4\times 10^{19}$~eV, where the GZK cutoff begins. 
The agreement of the dip with all data is very good. However, one can
see that the modification factor in the Akeno-AGASA data and in 
HiRes data exceeds 1 at $E \lsim 1\times 10^{18}$~eV. By definition 
the modification factor cannot be larger than 1. The excess of
modification factor over 1 signals the appearance of the new component
of  cosmic rays, which can be nothing but galactic cosmic rays. Thus ,
this excess evidences for transition from galactic to 
extragalactic cosmic  rays. Strictly speaking at the energy 
 $E \sim 1\times 10^{18}$~eV 

%\bigskip
\noindent
transition is completed and at higher
energy the cosmic rays are strongly dominated by extragalactic
component. The transition occurs at lower energy, at the second knee 
seen at the range $(4 - 8)\times 10^{18}$~ eV in the different experiments.
\begin{figure*}[t]
\begin{center}
\mbox{\includegraphics[width=0.65\textwidth,height=5.8cm]{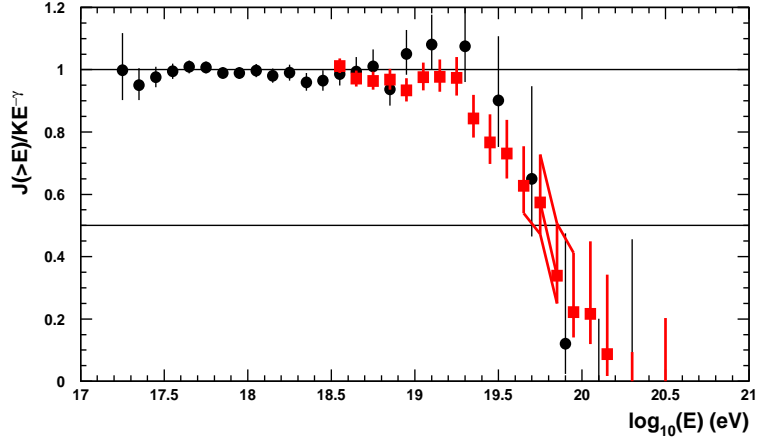}}
\end{center}
\caption{$E_{1/2}$ as numerical characteristic of the GZK cutoff in
  the integral HiRes spectrum.
}
\label{fig:E-half}
\end{figure*}
\noindent
The steepening observed in the  HiRes spectrum is confirmed as the GZK 
feature by the measured value of $E_{1/2}$ in the integral spectrum. 
$E_{1/2}$ is a characteristic of the GZK cutoff in the integral
spectrum \cite{BG88}. It is defined as energy where  
the integral spectrum calculated with all energy losses included
becomes half of the power-law extrapolation $KE^{-\gamma}$ from the low 
energies. In \cite{BG88} and \cite{BGG-prd} it was demonstrated that 
$E_{1/2}$ is a model-independent value which equals to
$10^{19.72}$~eV. Fig.~\ref{fig:E-half} shows how $E_{1/2}$ was
found from integral HiRes spectrum. The ratio $\kappa=J(>E)/KE^{-\gamma}$ 
is plotted as function of energy, where $J(>E)$  is measured integral
spectrum and $KE^{-\gamma}$  is its power-law extrapolation. The equality
$\kappa \approx 1$ shows that the chosen approximation is good.
The energy where integral spectrum crosses the line $\kappa=1/2$
gives $E_{1/2}$. The corresponding value is found \cite{E-half-Hi} 
$E_{1/2}= 10^{19.73 \pm 0.07}$ in excellent agreement with the
theoretical value. 

The confirmation of the pair-production dip and the GZK cutoff in 
observational data evidences that primary spectrum at 
$E \gsim 1\times 10^{18}$~eV is strongly dominated by protons.\\*[2mm]                            
%%%%%%%%%%%%%%%%%%%%%%%%%%%%%%%%%%%%%%%%%%%%%%%%%%%%%%%%%%%%%%%%%%%%
{\bf Neutrino fluxes in the dip model}

To calculate the neutrino flux produced by UHE protons it is enough 
to know the generation rate of UHE protons at each cosmological epoch, 
which we take in the form $Q(E)(1+z)^m$, where factor $(1+z)^m$
describes the cosmological evolution of the sources. One should also 
introduce $z_{\rm max}$ up to which 
the assumed evolution of the sources holds. 
$Q(E)$ gives the rate of proton generation at $z=0$, i.e. 
the number of protons with energy E generated per unit of comoving
volume per unit time. Expressed in terms of the Lorentz factor 
$\Gamma=E/m_p$ and the emissivity ${\mathcal L}_0$ at  $z=0$,
i.e. energy generated per unit comoving volume and unit time, the
generation rate is given by 
\be
Q(\Gamma)= (\gamma_g - 2) \frac{{\mathcal L}_0}{m_p}\Gamma^{-\gamma_g},
\label{Q}
\ee
where $\gamma_g = 2.6 - 2.7 $ is the generation index and 
$\Gamma_{\rm min} \sim 1$ is assumed.  

However, following the works \cite{KS,Aloisioetal} we assume that acceleration
index in each source is the same $\gamma_{\rm acc}=2.0 - 2.2$, but the maximum
energies of acceleration $E_{\rm acc}^{\rm max}$ are different and 
distribution of sources over $E_{\rm acc}^{\rm max}$ results in 
steepening of the generation spectrum at $E \geq E_c$ to 
$\gamma_g \approx 2.6 - 2.7$. 

\begin{figure*}[h]
  \begin{center}
  \mbox{\includegraphics[width=12cm]{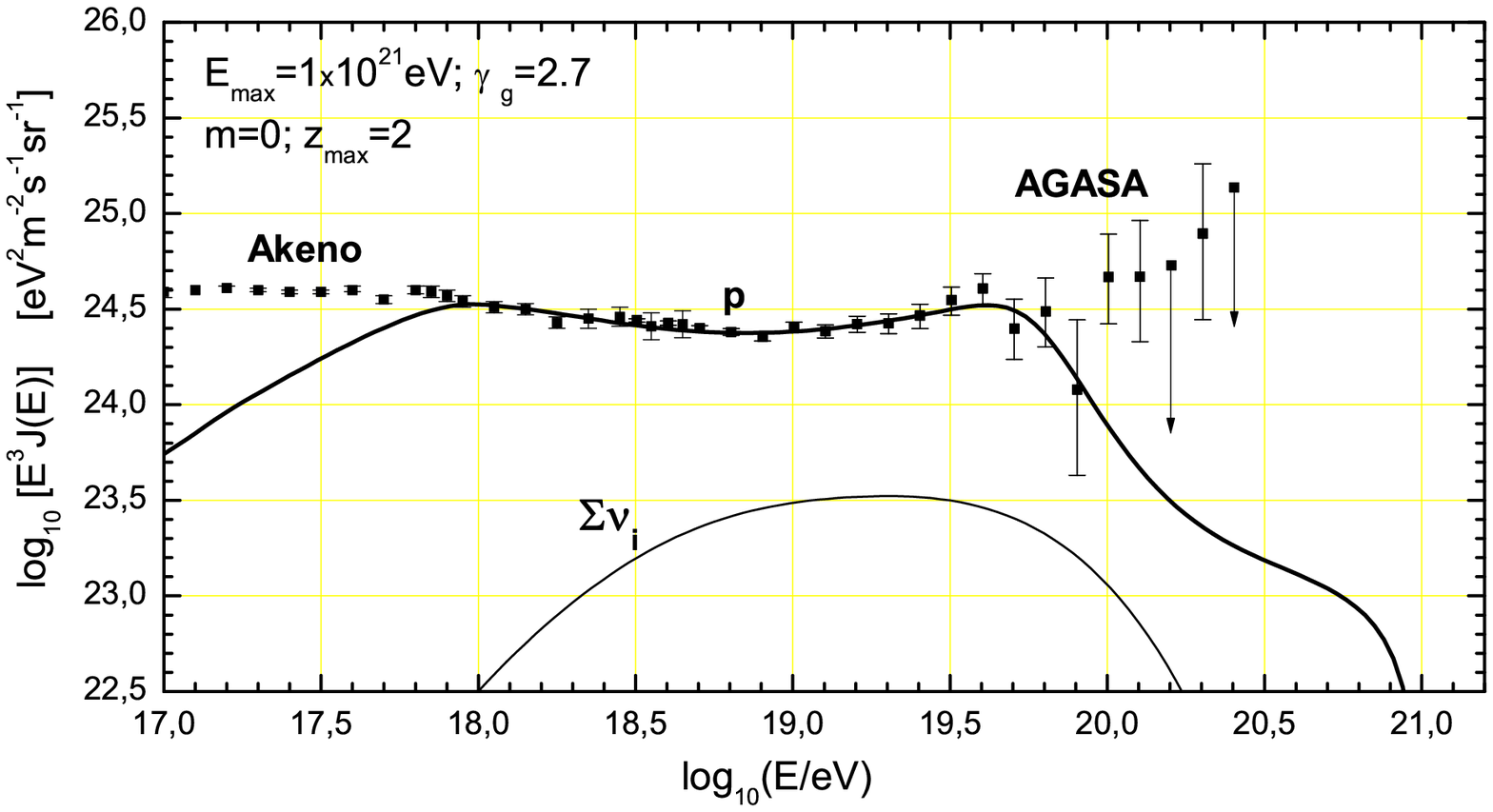}}
  \end{center}
\begin{center}
  \mbox{\includegraphics[width=12cm]{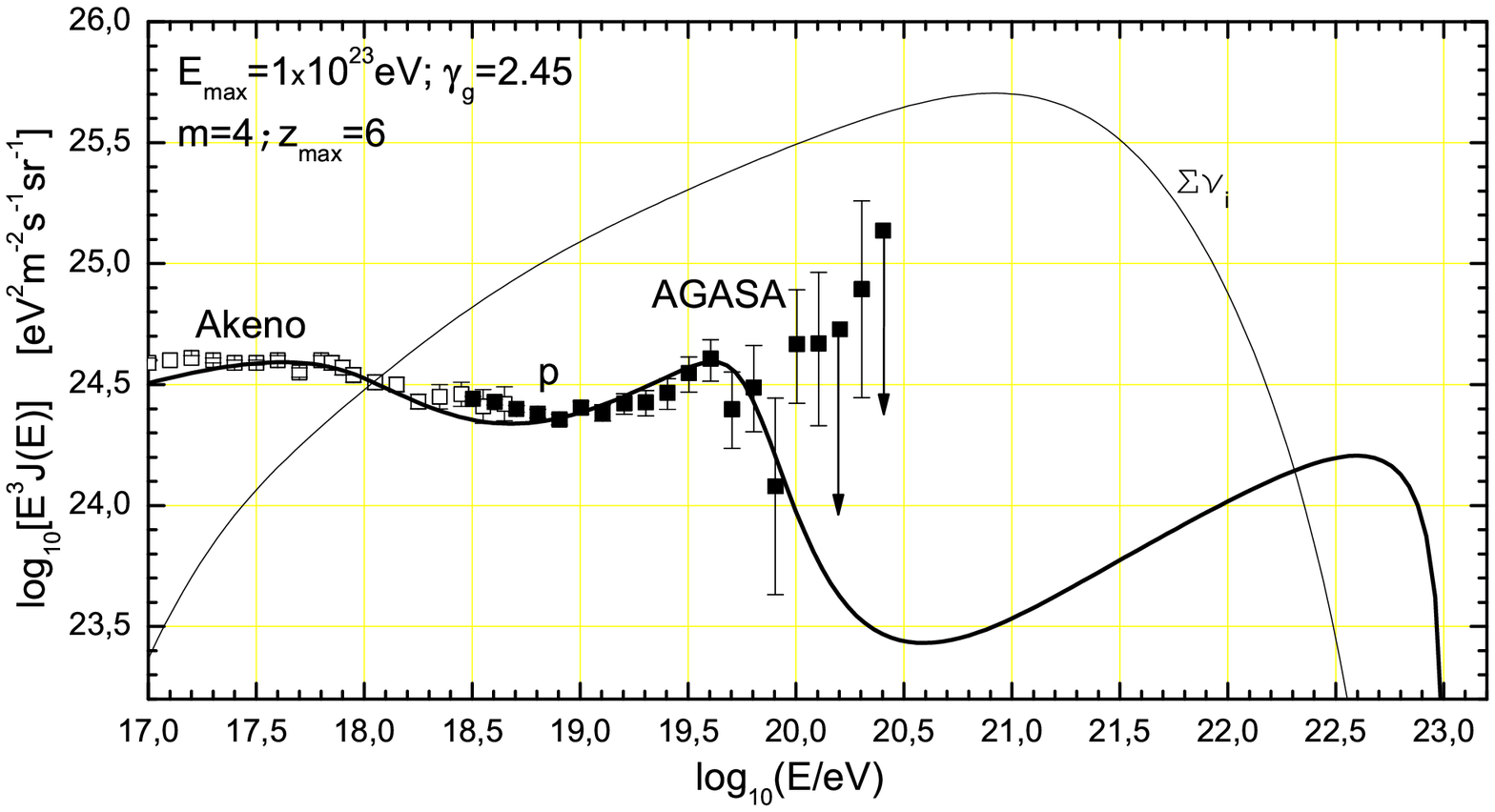}}
  \end{center}
\vspace*{-2mm}
\caption{UHE neutrino fluxes  in the non-evolutionary 
 dip model (upper panel) and in the  
 evolutionary dip model (lower panel).
 The neutrino fluxes are shown 
 by curves $\Sigma\nu_i$ for the sum of all  neutrino flavours. The  following 
  parameters are used in calculations: $m=0$,~ $z_{\rm max}=2$,~ 
  $\gamma_g=2.7$,~ $E_{\rm max}=1\times 10^{21}$~eV,~ 
  $E_c=1\times 10^{18}$~eV and
  ${\cal L}_0=3.5\times 10^{46}$~erg/Mpc$^3$yr (upper panel).  
  In the lower panel the neutrino flux is maximized by the following 
  choice of parameters: $m=4.0$, $z_{\rm max}=6.0$,~ $\gamma_g=2.45$ 
  and emissivity ${\cal L}_0= 1.2\times 10^{46}$~erg Mpc$^{-1}$yr$^{-1}$.
  }
 \label{fig:uhenu-dip}
\end{figure*}

The model with the {\em minimum UHE neutrino flux} corresponds to absence  
of the evolution $m=0$ and low maximum energy of acceleration 
$E_{\rm acc}^{\rm max}=1\times 10^{21}$~eV. Less important 
is the assumption for value of $E_c$, for which we use  
$E_c \sim 1\times 10^{18}$~eV, The calculated minimum flux is shown in 
the upper panel of Fig.~\ref{fig:uhenu-dip}. This flux is very small, 
but it could be marginally detectable by IceCube at 
$E \gsim 10^{17}$~eV and by JEM-EUSO at $E \gsim 10^{19}$~eV. 

\begin{figure*}[h]
\begin{center}
\mbox{\includegraphics[width=0.7\textwidth,height=6.1cm]{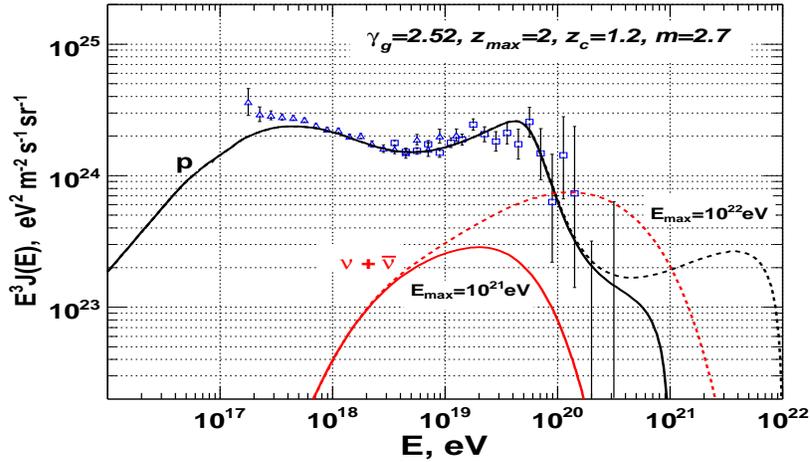}}
\end{center}
\caption{UHE neutrino flux in the dip model with AGN as the sources of 
UHECR. The cosmological evolution of AGN with $m=2.7$ up to 
$z_c = 1.2$  is taken from  X-ray observations of AGN. At larger $z$
the evolution is frozen up to $z_{\rm max}=2.0$. The fit of the dip 
is very good, though requires  $\gamma_g = 2.52$ different from the 
non-evolutionary case $m=0$. The neutrino fluxes are given for one
neutrino flavour and they are detectable especially for the case 
$E_{\rm acc}^{\rm max}=1\times 10^{22}$~eV. 
}
\label{fig:AGN}
\end{figure*}

One can maximize the flux introducing the cosmological evolution 
of the sources and assuming high maximum acceleration energy. 
This flux is shown in the lower panel of Fig.~\ref{fig:uhenu-dip} for
the following choice of parameters: $m=4.0$, $z_{\rm max}=6.0$,~ 
$\gamma_g=2.45$ and emissivity ${\cal L}_0= 
1.2\times 10^{46}$~erg Mpc$^{-1}$yr$^{-1}$.
One may notice the worse agreement with the
dip with than in the case of the non-evolutionary model.\\    
{\bf Neutrino fluxes from AGN in the dip model}\\
AGN are the most promising sources of the observed UHECR  
as far as acceleration and total energy output is concerned 
\cite{BGG-AGN,BGG-prd}. There are also some observational indications
in the form of the correlations of UHECR particles with AGN 
\cite{Tkachev,Farrar,AGN-Auger}. We present here the results of calculations 
\cite{BGG-nu} performed in phenomenological approach. We assume the
generation rate of UHE protons as described above with spectral index 
$\gamma_g= 2.0$ at $E \leq E_c$ and $\gamma_g = 2.52$ above this
energy due to assumed distribution of AGN over $E_{\rm max}$. The
value $\gamma_g = 2.52$ is chosen to provide the best fit to the dip 
(see Fig.~\ref{fig:AGN}). The UHE neutrino flux is calculated due to
interaction with CMB. The evolution of AGN is taken according to X-ray
observations \cite{AGN-evolution}: $(1+z)^m$ with $m=2.7$ up to 
$z_c=1.2$. At larger $z$ the evolution is frozen up to 
$z_{\rm max}=2$. Notice, that this is very weak evolution, but 
even such evolution combined with  $E_{\rm acc}^{\rm max}=1\times 10^{22}$~eV
makes UHE neutrino fluxes detectable by EUSO and radio
detectors. Neutrino fluxes in Fig.~\ref{fig:AGN} are given for one
neutrino flavour. 
%%%%%%%%%%%%%%%%%%%%%%%%%%%%%%%%%%%%%%%%%%%%%%%%%%%%%%%%%%%%%%%%%%%%%%
\section{Cascade upper limit on UHE neutrino flux}
The Waxman-Bahcall \cite{WB} upper bound is not applicable for 
UHE neutrinos. This bound is obtained from equality of the accompanying 
UHE proton flux and the observed flux of UHECR, assuming 
some relation between proton and neutrino fluxes at production. 
However, one can see 
from Fig.~\ref{fig:uhenu-dip} that UHE neutrino fluxes differ for the
same proton flux by two-three order of magnitudes at different
energies (see upper and lower panels). The most interesting case 
of UHE neutrino flux produced by top-down models are not limited by 
the Waxman-Bahcall bound because accompanying proton flux is negligibly
small and most produced particles are pions and kaons. 

However, the Waxman-Bahcall limit is useful as a low-flux 
benchmark for the future experiments. 

The most efficient upper bound for UHE neutrinos, applicable for both 
cosmogenic neutrinos and neutrinos from top-down models, is given by
the {\em cascade upper limit} first considered in \cite{BS} (see also 
\cite{pylos}).

\begin{figure*}[h]
  \begin{center}
  \mbox{\includegraphics[width=0.7\textwidth]{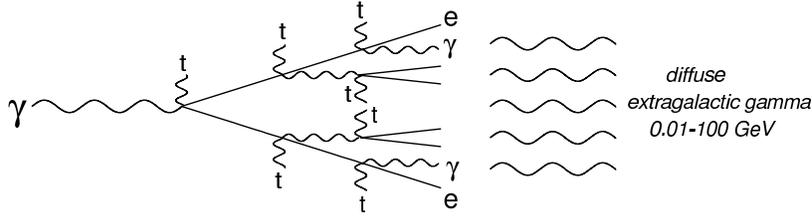}}
  \end{center}
  \caption{Electromagnetic cascade developing on the target photons (t), 
   e.g. CMB.   
}
\label{fig:cascade} 
  \end{figure*}

The cascade upper limit on UHE neutrino fluxes \cite{BS,book,nucl}
is provided due to e-m cascades initiated by UHE photons or electrons  
which always accompany production of UHE neutrinos. Colliding with 
low-energy target photons, a primary photon or electron produce 
e-m cascade due to reactions $\gamma+\gamma_{\rm tar} \to e^++e^-$,
$e+\gamma_{\rm tar} \to e'+\gamma'$, etc. (see Fig.~\ref{fig:cascade}).  
The standard case, valid for cosmogenic neutrinos, is given
by production of UHE neutrinos in extragalactic space, and the cascade 
develops due to collisions with CMB photons ($\gamma_{\rm tar}= 
\gamma_{\rm CMB}$). In case neutrino production occurs in a galaxy,
the accompanying photon can either freely escapes from a galaxy
and produce cascade in extragalactic space, or produce cascade on 
the background radiation (e.g. on CMB or infrared) inside the galaxy. In the 
latter case the galaxy should be transparent for the cascade photons 
in the range 10~ MeV - 100~GeV. 
The spectrum of the cascade photons is calculated 
\cite{BS,book,nucl,Blasi-cas}:
in low energy part it is $\propto E^{-3/2}$, at high energies  
$\propto E^{-2}$ with a cutoff at some energy $\epsilon_{\gamma}$.
The energy of transition between two regimes is given approximately by
$\epsilon_c \approx (\epsilon_t /3)(\epsilon_{\gamma} /m_e)^2$, 
where $\epsilon_t$ is the mean energy of the target photon. 
In case the cascade develops in extragalactic space
$\epsilon_t=6.35\times 10^{-4}$~eV,
$\epsilon_{\gamma} \sim 100$~GeV (absorption on optical radiation),
and $\epsilon_c \sim 8$~MeV. The cascade spectrum is very close 
to the EGRET observations in the range 3~MeV - 100~GeV \cite{EGRET}.  
The observed energy density in this range is 
$\omega_{\rm EGRET} \approx (2 - 3)\times 10^{-6}$~eV/cm$^3$.
It provides the upper limit for the cascade energy density.  
The upper limit on UHE neutrino flux $J_{\nu}(>E)$ (sum of all flavors) 
is given by chain of the following inequalities  
\be
\omega_{\rm cas}>\frac{4\pi}{c}\int_E^{\infty}EJ_{\nu}(E)dE>
\frac{4\pi}{c}E\int_E^{\infty}J_{\nu}(E)dE\equiv
\frac{4\pi}{c}EJ_{\nu}(>E),
\label{eq:int-limit}
\ee
which in terms of the differential neutrino spectrum $J_{\nu}(E)$ gives
\begin{equation}
E^2 J_{\nu}(E) < \frac{c}{4\pi}\omega_{\rm cas },~~ {\rm with}~
\omega_{\rm cas} <\omega_{\rm EGRET}
\label{cas-rig}
\end{equation}

\begin{figure*}[t]
\begin{center}
\mbox{\includegraphics[width=12cm,height=7cm]{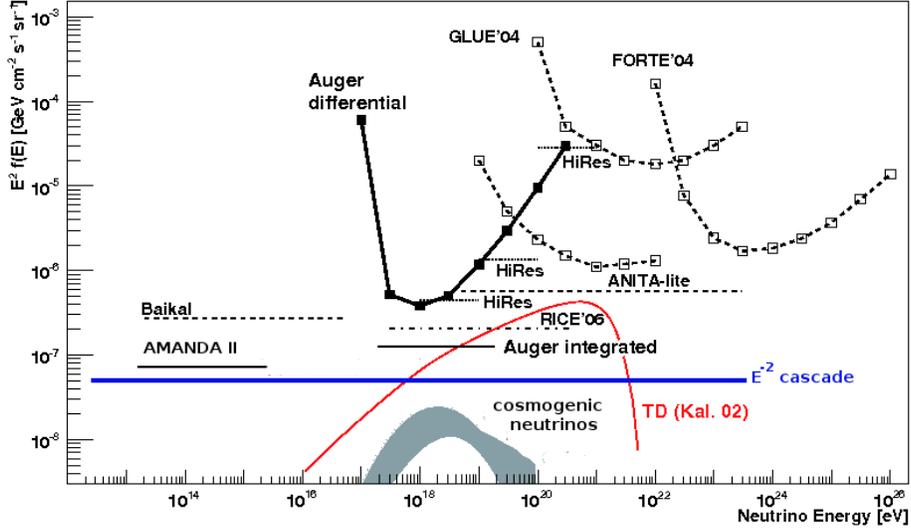}}
\end{center}
\caption{ The experimental upper limits on UHE neutrino fluxes 
in comparison with the e-m cascade upper limit in assumption of 
$E^{-2}$ generation spectrum (curve $E^{-2}$ cascade) and with predictions 
for cosmogenic neutrinos and neutrinos from TDs \protect\cite{Kal}. 
The plot is the modified one from \protect\cite{Kampert}. 
}
\label{fig:upper-limits}
\end{figure*}
Eq.~(\ref{cas-rig}) gives the {\em rigorous} upper limit on the neutrino flux. 
It is valid for neutrinos produced by HE protons, by topological defects, by
annihilation and decays of superheavy particles, i.e. in all cases
when neutrinos are produced through decay of pions and kaons. It is
valid for production of neutrinos in extragalactic space and in
galaxies, if they are transparent for the cascade photons. It holds
for arbitrary neutrino spectrum falling down with energy. If one assumes 
some specific shape of neutrino spectrum, the cascade limit becomes stronger.   
For $E^{-2}$ generation spectrum, which is usually assumed for
analysis of observational data  one obtains the stronger upper
limit.  Given for one neutrino flavour it reads  
\begin{equation}
E^2J_i(E) \leq \frac{1}{3} \frac{c}{4\pi}\frac{\omega_{\rm cas}}
{\ln (E_{\rm max}/E_{\rm min})},
\label{cas-E2}
\end{equation}
where $i=\nu_{\mu}+\bar{\nu}_{\mu}$ or $i=\nu_e+\bar{\nu}_e$.\\
This upper limit is shown in Fig.~\ref{fig:upper-limits}.
One can see that the observations almost reach the cascade upper 
limit and thus almost enter the region of allowed fluxes. 

The most interesting energy range in Fig.~\ref{fig:upper-limits} 
corresponds to 
$E_{\nu} > 10^{21}$~eV, where acceleration cannot provide protons 
with sufficient energy for production of these neutrinos. In fact 
this statement is valid only for shock acceleration. In principle, 
e.g. in AGN the different mechanisms of acceleration might operate,
such as unipolar induction or pinch-instability, and they can provide
the higher energy of acceleration, but these mechanisms are not
developed enough for numerical calculations of produced fluxes of 
accelerated particles. 
At present the region of $E_{\nu} > 10^{21}$~eV, and especially   
$E_{\nu} \gg 10^{21}$~eV is considered as a signature of 
top-down models, which provide these energies quite naturally. 
Below we shall consider three top-down models: Superheavy Dark Matter,
Topological Defects and Mirror Matter. 

%%%%%%%%%%%%%%%%%%%%%%%%%%%%%%%%%%%%%%%%%%%%%%%%%%%%%%%%%%%%%%%%%%%
\section{UHE neutrinos from Superheavy Dark Matter (SHDM)}  
SHDM is one of the models for cosmological cold dark matter 
\cite{BKV,whimpzillas}. The most attractive mechanism of
production is given by creation of superheavy particles in
time-varying gravitational field in post-inflation 
epoch \cite{Kolb-grav,Kuz-grav}. Creation 
occurs when the Hubble parameter is of order of particle mass 
$H(t) \sim m_X$. Since the maximum value of the Hubble parameter is limited  
by the mass of the inflaton $H(t) \lsim m_{\phi} \sim 10^{13}$~GeV,  
the mass of X-particle is limited by $m_{\phi}$, too. For example, 
$m_X \sim 3\times 10^{13}$~GeV results in $\Omega_X h^2 \sim 0.1$, as
required by WMAP measurements. 

Being protected by some symmetry, SHDM particles with such masses can be 
stable or quasi-stable. In case of gauge symmetry they are stable, in
case of gauge discrete symmetry they can be stable or quasi-stable. 
Decay can be provided by superweak effects: wormholes, instantons, 
high-dimension operators etc. 

Like any other form of cold dark matter, X-particles are accumulated
in the halo with overdensity $2.1\times 10^{5}$. 

SHDM particles can produce UHECR and high energy neutrinos at the
decay of X-particles (when the protecting symmetry is broken) and at
their annihilation, when the symmetry is exact. The scenario with 
decaying X-particles was first studied in \cite{BKV,KR,Sarkar}.
An interesting scenario with stable X-particles, when UHE particles
are produced by annihilation of X-particles has been put forward in 
\cite{Khlopov}. 
In this scenario superheavy X-particles have the gauge charge and they
are produced at post-inflationary epoch by close pairs, forming the 
bound systems. Loosing the
angular momentum, these particles inevitably annihilate in a close
pair.   

\begin{figure*}[h]
  \begin{center}
  \mbox{\includegraphics[width=0.7\textwidth]{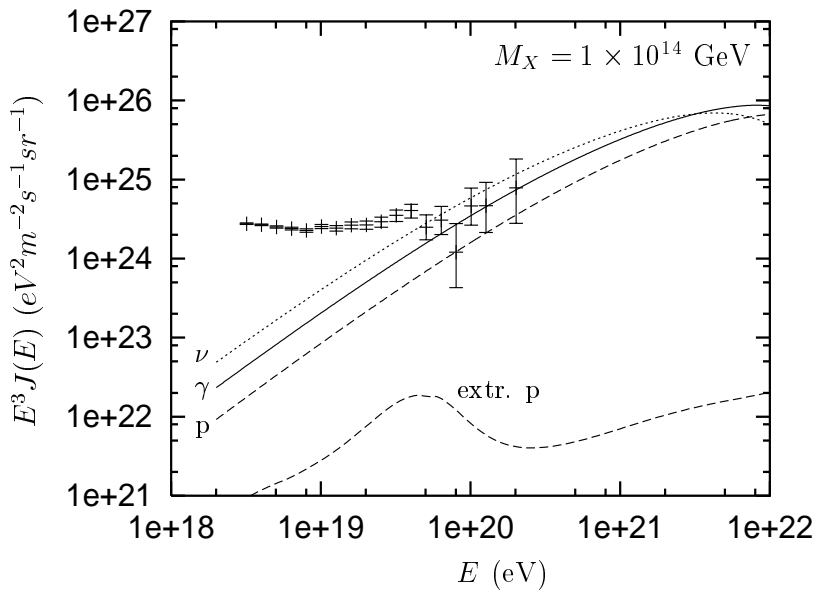}}
  \end{center}
  \caption{Spectra of neutrinos (upper curve), photons (middle curve)
    and protons (two lower curves) in SHDM model compared with AGASA
    data, according to calculations of \protect\cite{DGLAP-ABK}.
    The neutrino flux is dominated by the halo
    component with small admixture of extragalactic flux. The flux of
    extragalactic protons is shown by the lower curve (extr. p).
    The fluxes are normalized by the AGASA excess at 
    $E \gsim 1\times 10^{20}$~eV. In case it is absent, like in HiRes
    and Auger data, all fluxes, including neutrino, must be lowered by
    factor 3 - 5.}
    \label{SHDM}
  \end{figure*}

The UHE particles (protons, pions and neutrinos from 
the chain of pion decays)  are produced as a result of QCD cascading
of partons. The calculations of fluxes and spectra are nowadays
reliably performed by Monte Carlo \cite{BK-MC} and using the  
DGLAP equations \cite{DGLAP-Sarkar} - \cite{DGLAP-ABK}.  
The spectra of protons, photons and 
neutrinos are shown in Fig.~\ref{SHDM} for the case of SHDM particles 
with mass $M_X=1\times 10^{14}$~GeV. The spectrum of photons is
normalized by the AGASA excess. In case it is absent, like in HiRes
and Auger observations, all fluxes, including neutrino, must be  
lowered by factor 3 - 5.   

\section{UHE neutrinos from Topological Defects (TDs)}
As has been first noticed by D.~A.~Kirzhnitz \cite{Kirzhnitz},
each spontaneous symmetry breaking in the early universe is accompanied by 
the phase transition. Like the phase transitions in liquids and
solids, the cosmological phase transitions can give rise to
topological defects (TDs), which can be in the form of surfaces
(cosmic textures), lines (cosmic strings)
and points (monopoles). 
In many cases TDs become unstable and decompose to constituent fields, 
superheavy gauge and Higgs bosons (X-particles), which then decay 
producing UHECR. It could happen, for example, when two segments of 
ordinary string, or monopole and antimonopole touch each other, when 
electrical current in superconducting string reaches the critical value
and in some other cases. The decays of these particles, if they heavy enough, 
produce particles of ultrahigh energies including neutrinos.

The following TDs are of interest for UHECR and neutrinos \cite{BBV}:\\  
{\em monopoles} ($G \to H\times U(1)$ symmetry breaking), 
{\em ordinary strings}
($U(1)$ symmetry breaking) with important subclass of 
superconducting strings,  {\em monopoles connected by strings} 
($G \to H\times U(1)$ symmetry breaking with subsequent $U(1) \to Z_N$
symmetry breaking, where $Z_N$ is discrete symmetry). The important
subclass of the monopole-string network is given by {\em necklaces},
when $Z_N=Z_2$, i.e. each monopole is attached to two strings. 
We shall shortly describe the production of UHE particles by these TDs.
(see \cite{BBV} for more details).

(i) {\em Ordinary strings}.\\
There are several mechanisms by which ordinary strings can produce UHE 
particles.
For a special choice of initial conditions, an ordinary  string loop
can collapse to a
double line, releasing its total energy in the form of X-particles. 
However, the probability of this mode of collapse is
extremely small, and its contribution to the overall flux of UHE
particles is negligible.

String loops can also 
produce X-particles when they self-intersect.
Each intersection, however, gives only a few
particles, and the corresponding flux is very small. 

The loops undergo oscillation and an important property of it 
is the periodic appearance of the cusps, the loop points  with
velocity of light.  
The near-cusp segments, moving with large Lorentz factor may overlap
and annihilate producing the constituent superheavy particles with 
large Lorentz factors. Therefore the particles from X-decays are
further boosted by large Lorentz factor.  
The energy released in a single cusp event can be quite large, but
again, the resulting flux of UHE particles is too small to account for
the UHECR observations.

(ii) {\em Superconducting strings}.\\
As was first noted by Witten\cite{Witten}, in a wide class of elementary 
particle models, strings behave like superconducting wires. Moving through 
cosmic magnetic fields, such strings develop electric currents.
Superconducting strings produce X-particles when the electric current
in the strings reaches the critical value. This process is strongly 
increased in cusps where in a small fraction of loop the current
becomes supercritical and X-particles, the charge carriers, leave 
the string and are decaying to the high-energy ordinary particles.  
Their energies are further boosted by cusp Lorentz factor. 
Superconducting strings
cannot  be the sources of observed UHECR \cite{BBV} because of the
absorption of the particles on the way from a source to observer, but
they can produce the observable flux of UHE neutrinos. 

(iii){\em Network of  monopoles connected by strings}.\\
The sequence of phase transitions
\begin{equation}
G\to H\times U(1)\to H\times Z_N
\label{symm}
\end{equation}
 results in the formation of monopole-string networks in which each monopole 
is attached to N strings. Most of the monopoles and most of the strings belong 
to one infinite network. The evolution of networks is expected to be 
scale-invariant with a characteristic distance between monopoles 
$d=\kappa t$, where $t$ is the age of Universe and $\kappa=const$. 
The production of UHE particles are considered in \cite{BMV}. Each 
string attached 
to a monopole pulls it with a force equal to the string tension, $\mu \sim 
\eta_s^2$, where $\eta_s$ is the symmetry breaking vev of strings. Then
monopoles have a typical acceleration $a\sim \mu/m$, energy $E \sim \mu d$ 
and Lorentz factor $\Gamma_m \sim \mu d/m $, where $m$ is the mass of the 
monopole. Monopole moving with acceleration can, in principle, radiate  
gauge quanta, such as photons, gluons and weak gauge bosons, if the
mass of gauge quantum (or the virtuality $Q^2$ in the case of gluon) is
smaller than the monopole acceleration. The typical energy of radiated quanta 
in this case is $\epsilon \sim \Gamma_m a$. This energy can be much higher 
than what 
is observed in UHECR. However, the produced flux (see \cite{BBV}) is much 
smaller than the observed one. 

(vi){\em Necklaces}.\\
Necklaces are hybrid TDs corresponding to the case $N=2$ , i.e. to the
case when each monopole is attached to two
strings.  This system resembles ``ordinary'' cosmic strings,
except the strings look like necklaces with monopoles playing the role
of beads. The evolution of necklaces depends strongly on the parameter
\begin{equation}
r=m/\mu d,
\end{equation}
where $m$ is a mass of a monopole, $\mu$ is mass per unit length of a 
string (tension of a string) and 
$d$ is the average separation between monopoles and antimonopoles
along  the strings.
As it is argued in Ref.~\cite{BV}, necklaces might evolve to  
configurations with $r\gg 1$.  
Monopoles and antimonopoles trapped in the necklaces
inevitably  annihilate in the end, producing first the heavy  Higgs and 
gauge bosons ($X$-particles) and then hadrons.
The rate of $X$-particle production can be estimated as \cite{BV} 
\begin{equation}
\dot{n}_X \sim \frac{r^2\mu}{t^3m_X}.
\label{xrate}
\end{equation}
This rate determines the rates of pion and neutrino production
with energy spectrum calculated in Ref.~\cite{DGLAP-ABK}. 
\begin{figure*}[ht]
  \begin{center}
  \mbox{\includegraphics[width=0.7\textwidth]{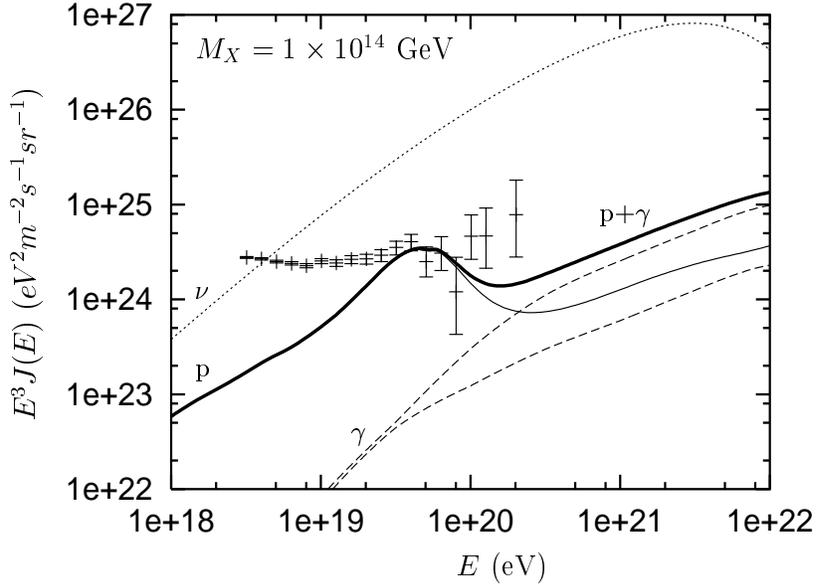}}
  \end{center}
  \caption{Diffuse spectra of neutrinos, protons and photons from
  necklaces. The upper curve shows neutrino flux, the middle - proton
  flux and two lower curves - photon fluxes for two cases of
  absorption. The thick curve gives the sum of the proton and the higher 
  photon flux.}
  \label{neckl}
  \end{figure*}
Restriction due to e-m cascade radiation demands the cascade energy density 
$\omega_{cas} \leq 2\cdot 10^{-6}$~eV/cm$^3$. The cascade energy density 
produced by necklaces can be calculated as
\begin{equation}
\omega_{cas}=
\frac{1}{2}f_{\pi}r^2\mu \int_0 ^{t_0}\frac{dt}{t^3}
\frac{1}{(1+z)^4}=\frac{3}{4}f_{\pi}r^2\frac{\mu}{t_0^2},
\label{eq:n-cas}
\end{equation}
where $f_{\pi}\approx 0.5$ is a fraction of total energy release 
transferred to the cascade. Therefore, $r^2\mu$ 
and the rate of X-particle production (\ref{xrate}) is limited by cascade 
radiation.

The fluxes of UHE protons, photons and neutrinos from necklaces  are shown in 
Fig.~\ref{neckl} according to 
calculations of \cite{DGLAP-ABK}. The mass of X-particle
is taken $m_X=1\times 10^{14}$~GeV. Neutrino flux is noticeably higher
than in the case of conservative scenarios for  cosmogenic neutrinos
and neutrinos from SHDM.

In the recent work \cite{Olum} it was indicated that in some models 
of necklaces there could be fast annihilation of monopoles and 
separation of pairs monopole-antimonopole connected by strings from
the necklaces. In these models UHE neutrino flux is suppressed. 
However, in some other models the detectable neutrino flux can exist. 

\section{Mirror matter and mirror neutrinos}
\label{sec:mirror}
Existence of mirror matter in our universe is the old idea which was 
put forward in the end of 1950s. 
Mirror matter can be most powerful source of UHE neutrinos not limited
by e-m cascade limit. Produced as the mirror neutrinos, they can
oscillate into ordinary neutrinos. All mirror particles that accompany 
production of mirror neutrinos in the mirror matter remain invisible 
for our detectors.\\*[2mm] 
{\bf Concept of mirror matter}\\*[1mm]
Mirror matter is based on the theoretical concept of 
the space reflection, as first suggested by Lee and Yang \cite{LY} in 1956
and developed by Landau \cite{Landau} in 1957, Salam \cite{Salam} in
1957, and most notably by Kobzarev, Okun, Pomeranchuk \cite{KOP} in
1966 (see recent exciting historical review by Okun \cite{Okun06}). 

This concept can be explained in the following way.

The Hilbert particle space is assumed to be a representation of 
the extended Lorentz group, which includes the 
space coordinate reflection $\vec{x} \to -\vec{x}$. Since the coordinate 
operations, reflection $\vec{x} \to -\vec{x}$ and time shift 
$t \to t+\Delta t$,  commute, the corresponding
operations in the particle space, $I_r$ and 
Hamiltonian $H$, must commute, too:  
\be
\left [{\mathcal H},I_r \right]=0
\label{eq:commutation}
\ee
It implies that operator $I_r$ must correspond to the conserved value.  
Since parity $P$ according to assumption of Lee and Yang is not conserved,
$I_r$ should be defined somehow else. Lee and Yang suggested 
$I_r=P \times R$, where $R$ transfers particle to mirror particle. 
Since parity operator $P$ interchange left and right states, one obtains   
\be 
I_r \Psi_{\rm L}=\Psi'_{\rm R}~~~{\rm and}~~~ I_r \Psi_{\rm R}=\Psi'_{\rm L},
\label{eq:transform}
\ee
where primes indicate the states in mirror particle space. 

The assumption of Landau \cite{Landau} was 
$R=C$, i.e. one may say that he suggested to use antiparticles 
as the mirror space, but then $CP$ must be conserved
which as we know today is not the case.\\*[2mm] 
{\bf Oscillation of mirror and ordinary neutrinos}\\*[1mm]
Kobzarev,~ Okun,~ and Pomeranchuk, \cite{KOP} suggested that ordinary
and mirror sectors communicate only gravitationally. 
For the description of this interaction one can
use dimension 5 operator obtained as 
$SU(2)_L\times U(1)\times SU(2)'_R\times U(1)'$ scalar: 
\be 
{\mathcal L_{\rm comm}} = \frac{1}{M_{\rm Pl}}
(\nu_{\rm L}\phi)(\nu'_{\rm R}\phi'),
\label{eq:comm}
\ee
where $M_{\rm Pl}= 1.2\times 10^{19}$~GeV is the Planckian mass,
which implies the gravitational interaction, 
and $\phi, \phi'$ are the electroweak Higgses 
from visible and mirror sectors, 

After spontaneous electroweak symmetry breaking 
$\langle \phi \rangle = \langle \phi' \rangle =v$
the Lagrangian (\ref{eq:comm}) 
generates the terms, which mix visible and sterile neutrinos,
\begin{equation}
{{\mathcal L}_{\rm mix}}=
\frac{v^2}{M_{\rm Pl}}\nu\nu',
\label{eq:mix}
\end{equation}
where $v$= 174 GeV is vacuum expectation value of Higgses,
and $\mu=v^2/M_{\rm Pl}=2.5\times 10^{-6}$~eV is the mixing parameter.

Eq.~(\ref{eq:mix}) implies oscillation of mirror and ordinary
neutrinos. This oscillation described by  Eq.~(\ref{eq:mix}), 
but with $M$ not necessarily being the Planckian mass, was first suggested 
by Berezhiani and Mohapatra \cite{BM} and Foot and Volkas \cite{FV} 
in 1995. \\*[2mm] 
{\bf UHE neutrinos from mirror TDs}\\*[1mm]
Any  cosmological scenario for mixed ordinary and mirror matter must 
provide the suppression of the mirror
matter and in particular the density of mirror photons and neutrinos 
at the epoch 
of nucleosynthesis. It can be obtained in the two-inflaton model 
\cite{mirror}. The rolling of two inlatons to minimum of the potential
is not synchronized, and when the mirror inflaton reaches minimum, the ordinary
inflaton continues its rolling, inflating thus the mirror matter produced
by the mirror inflaton. While mirror matter density is suppressed, the mirror
topological defects in two-inflatons scenario with curvature-driven
phase transition  can strongly dominate \cite{mirror}. Mirror TDs 
copiously produce mirror neutrinos with extremely high
energies typical for TDs, and they are not accompanied by
any visible particles. Therefore, the upper limits on HE mirror 
neutrinos in our world do not exist. All HE mirror particles 
produced by mirror TDs are sterile for us, interacting with
ordinary matter only gravitationally, and only mirror neutrinos 
can be efficiently converted into ordinary ones due to oscillations. 
The oscillations of mirror and visible neutrinos in the 
gravitational-mixing scenario has been studied in detail in 
\cite{BNV}. The probability of oscillation of mirror neutrino 
$\nu'$  into visible neutrinos is large. In particular, for 
$\nu'_{\mu}$ neutrino it is given by 
\be
P_{\nu'_{\mu} \nu_e}=\frac{1}{8}\sin^2 2\theta_{12}\;,\;\;
P_{\nu'_{\mu} \nu_{\mu}}=P_{\nu'_{\mu} \nu_{\tau}}= \frac{1}{4} - 
\frac{1}{6} \sin^2 2\theta_{12}\;,\;\; 
\sum_{\alpha} P_{\nu'_{\mu} \nu_{\alpha}}=\frac{1}{2}. 
\ee

%%%%%%%%%%%%%%%%%%%%%%%%%%%%%%%%%%%%%%%%%%%%%%%%%%%%%%%%%%%%%%%%%%%%%
\section{Conclusions}
UHE neutrino astronomy is characterised by well balanced program. 

As the secure part of this program, there are  cosmogenic
neutrinos, for production of which the particle beam (UHECR particles) 
and the target (CMB photons) are well known. In the case when UHECR 
primaries are protons and the observed dip is a feature of UHE proton
interaction with CMB, there is the robustly predicted {\em lower limit}
for UHE neutrino flux (see upper panel of Fig.~\ref{fig:uhenu-dip}), which
is marginally detectable by IceCube at $E < 1\times 10^{18}$~eV and by 
JEM-EUSO at $E > 1\times 10^{19}$~eV. This lower limit is supported
by the following observations: the features of UHE proton interaction      
with CMB, GZK cutoff and dip, are found, mostly in HiRes
observations, and the fluorescent data of HiRes evidence for
proton-dominated mass composition. However, the Auger fluorescent data
favour the mixed nuclei composition. In this case the cosmogenic
neutrinos can be undetectable for existing projects. 

The large fluxes of cosmogenic neutrinos, detectable by existing
projects, correspond (at the fixed UHECR flux) to cosmological 
evolution of the sources, flat generation spectra and large maximum 
energy of acceleration. 
The lower panel of Fig.~\ref{fig:uhenu-dip} presents UHE neutrino fluxes 
for the extreme hypothetical assumptions: very large $E_{\rm max}$ and
strong evolution of the sources up to $z_{\rm max} = 6$. 

The fundamental problem of astrophysics involved in prediction 
of UHE cosmogenic neutrinos is acceleration of particles. The shock 
acceleration at present knowledge of its theory cannot provide maximum
energy of acceleration higher than $10^{21} - 10^{22}$~eV, and thus
energies of cosmogenic neutrinos do not exceed $3\times 10^{20}$~eV. 
However, in case of AGN as UHECR sources, the alternative mechanisms 
of acceleration, such as unipolar induction and plasma mechanisms, 
can work and accelerate particles to much higher energies. Unfortunately,  
these mechanisms are not developed enough for numerical calculations 
of produced rates and spectra at generation. The detection of UHE
neutrinos from the individual sources can help to solve this
fundamental problem of astrophysics. 

The second part of UHE neutrino astronomy is provided by top-down 
sources: SHDM and topological defects in ordinary and mirror matter.  
The decays of unstable constituent fields of TDs can produce neutrinos 
with extremely high energies up to GUT scale and above. This is the
general property of top-down models. However, the fluxes are very 
model dependent, and in case of TDs they differ much for
different types of topological defects. There could be no flux
estimates based on the general common properties of all TDs: some of them 
give low undetectable fluxes, from some exceptional TDs one can
expect the detectable fluxes. For long time the necklaces have been  
considered as  most promising TDs for UHECR and neutrinos \cite{BV}, 
but recent work \cite{Olum} puts the doubts on these topological defects as 
sources of UHE particles in recent cosmological epochs. The signature 
of top-down models is very high neutrino energies, unreachable for 
accelerator neutrinos. These neutrinos are most reliably detectable  
by radio and acoustic methods, as well as by EUSO detectors. 

In both cases of cosmogenic and top-down models, the neutrino fluxes
are constrained by the cascade upper limit. 
As exception, the mirror neutrinos do not respect this  limit, and their 
fluxes can be even larger.

The search for UHE neutrinos in any case is a search for a new 
physics, either for astrophysics (the new acceleration mechanisms 
and cosmological evolution of the sources) 
or for topological defects, mirror topological defects and superheavy
dark matter.

\section{Acknowledgments}
 
I am grateful to my collaborators Roberto Aloisio, Pasquale Blasi, 
Askhat Gazizov, Svetlana Grigorieva, Alex Vilenkin and Francesco
Vissani for joint work and many useful discussions. This work is
partially supported by contract ASI-INAF I/088/06/0 for theoretical
studies in High Energy Astrophysics.

\end{document}